\documentclass{evn2024}
\usepackage{graphics,graphicx}
\usepackage{comment}
\usepackage{amsmath}

\begin{document}
   \title{High Resolution VLBI Imaging of Nearby Low Luminosity AGN jets}
   
   \author{Xi Yan\inst{1,2}
          \and
          Ru-Sen Lu\inst{1,3,4}
          }

   \institute{
    Shanghai Astronomical Observatory, Chinese Academy of Sciences, 80 Nandan Road, Shanghai 200030, China
    \and 
    Xinjiang Astronomical Observatory, Chinese Academy of Sciences, 150 Science-1 Street, Urumqi 830011, China
    \and 
    Key Laboratory of Radio Astronomy and Technology, Chinese Academy of Sciences, A20 Datun Road, Chaoyang District, Beijing, 100101, China
    \and
    Max-Planck-Institut f\"ur Radioastronomie, Auf dem Hügel 69, D-53121 Bonn, Germany}

   \abstract{
    Low-luminosity Active Galactic Nuclei (LLAGN) represent a unique class of AGN in the local universe. Extensive studies of these objects are essential for a comprehensive understanding of jet physics, as past research has largely focused on more powerful radio sources. In this report, we present our recent VLBI studies of two prominent nearby LLAGN, NGC 4261 and M104 (the Sombrero galaxy). Specifically, we address the kinematics, collimation, and fundamental physical parameters of their jets, and probe the possible origin of the radio emission at millimeter wavelengths.}
   \maketitle

\section{Introduction}
Very Long Baseline Interferometry (VLBI) is a unique tool for studying the physical processes at work in the Acceleration and Collimation Zones (ACZs) of the relativistic jets in AGNs. However, previous VLBI studies have been biased towards bright and powerful sources, such as blazars, which has two major drawbacks: 1) these high-luminosity sources are typically located at greater distances, and 2) the Doppler boosting due to their small viewing angles makes it more difficult to study the intrinsic properties of jets. As a result, our knowledge of jet physics remains very incomplete. With recent improvements in the sensitivity of various VLBI arrays, we now have the opportunity to fill this gap by studying weaker, but much closer sources: LLAGNs. In this report, we summarize our recent studies of the ACZs in two prominent LLAGNs, NGC\,4261 and M\,104, using high-resolution VLBI observations (see also Yan et al. \cite{Yan2023,Yan2024}). Table\,\ref{tab:tab_1} lists the masses of the supermassive black holes and the distances of NGC\,4261 and M\,104.

\vspace{-0.4cm} 
\begin{table}[htbp!]
	\begin{center}
	\caption{\centering Basic Information of NGC\,4261 and M\,104.}
        \vspace{-15pt}
	\label{tab:tab_1}
        \small
	\begin{tabular}{cccccccc} 
		\hline
		\hline
		LLAGN & $M_{\rm SMBH}$ & $D$ & Scale \\
              & ($10^9~M_{\odot}$) & (Mpc) & (mas$^{-1}$) \\
		\hline
            NGC\,4261 & $1.62\pm0.04$ & $31.6\pm0.2$ & $0.15\,\rm pc$ or $988\,R_{s}$\\
            M\,104 & $1.0$  & $9.55\pm0.31$ & $0.046\,{\rm\,pc} $ or $ 500\,R_{s}$ \\
		\hline
	\end{tabular}
	\end{center}
\end{table}

\vspace{-1.3cm} 
\section{Observations and Data Reduction}
\begin{figure*}[htbp!]
\begin{center}
\includegraphics[width=0.24\textwidth]{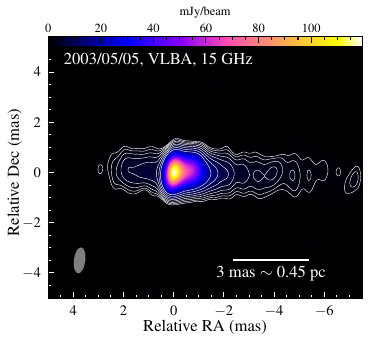}
\hspace{-3mm}
\includegraphics[width=0.265\linewidth]{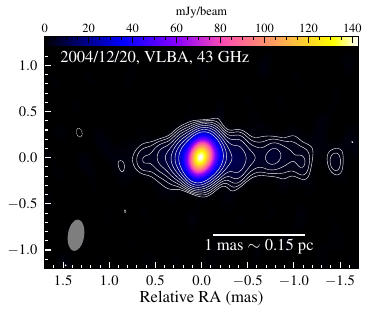}
\hspace{-3mm}
\includegraphics[width=0.26\linewidth]{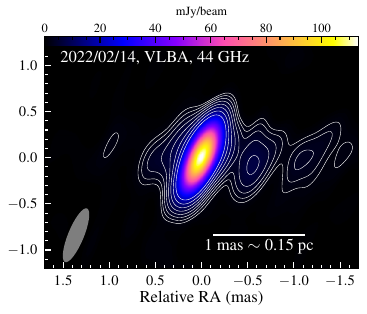}
\hspace{-3mm}
\includegraphics[width=0.245\linewidth]{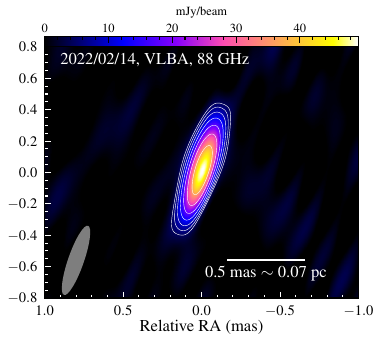}
\caption{Uniformly weighted CLEAN images of NGC\,4261 (Yan et al. \cite{Yan2023}).
\label{fig:NGC4261 maps}}
\end{center}
\end{figure*}

\begin{figure*}
\vspace{-0.3cm}
\begin{center}
\includegraphics[width=0.23\textwidth]{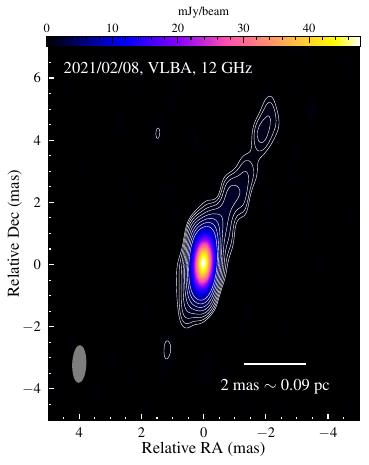}
\hspace{0.5cm}
\includegraphics[width=0.24\linewidth]{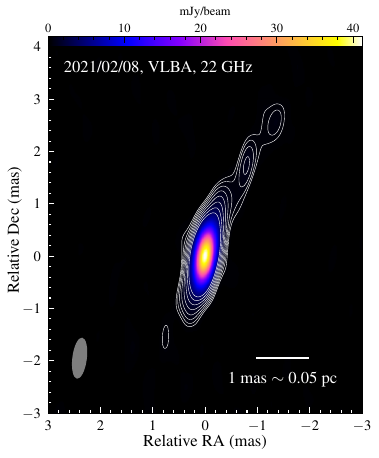}
\hspace{0.23cm}
\includegraphics[width=0.275\linewidth]{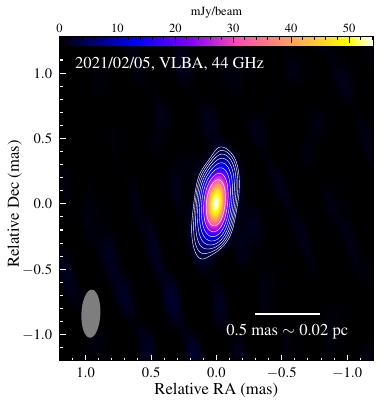}
\caption{Uniformly weighted CLEAN images of M\,104 (Yan et al. \cite{Yan2024}).
\label{fig:M104 maps}}
\end{center}
\end{figure*}

Our studies covered a wide frequency range between 1.4 and 88\,GHz, including both new and archival Very Long Baseline Array (VLBA) data. In particular, we performed Source-Frequency-Phase-Referencing (SFPR; Rioja et al. \cite{Rioja2011}) observations at 44 and 88\,GHz. The data were calibrated using AIPS (Greisen \cite{Greisen2003}), followed by self-calibration in DIFMAP (Shepherd \cite{Shepherd1997}). A comprehensive summary of these observations and detailed data reduction procedures can be found in Yan et al. (\cite{Yan2023}) and Yan et al. (\cite{Yan2024}). Example CLEAN images of NGC\,4261 and M\,104 are shown in Figs.\ref{fig:NGC4261 maps} and \ref{fig:M104 maps}, respectively. 

Notably, we successfully detected the nuclear structure of NGC\,4261 (with a total flux density of $66.6\pm6.7\,\text{mJy}$) using of SFPR technique. We also derived an upper limit on its core size ($0.09\pm0.02\,\text{mas}$). However, the imaging of M\,104 at 88\,GHz was difficult due to the very limited $u$-$v$ coverage. Therefore, we estimated a core size of $0.05\pm0.01$\,mas and a flux density of $61.9\pm11.2\,\text{mJy}$ directly from the visibility data (see Yan et al. \cite{Yan2024} for more details).

\begin{figure*}
\vspace{-0.3cm}
\begin{center}
        \includegraphics[width=0.4\textwidth]{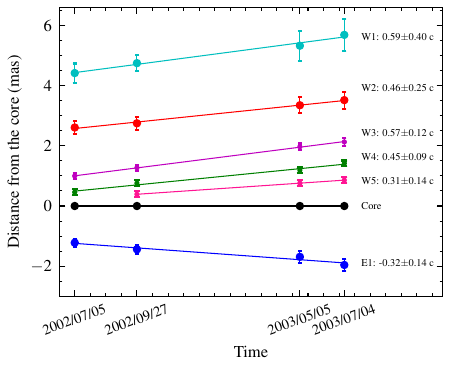}
        \hspace{5mm}
        \includegraphics[width=0.44\textwidth]{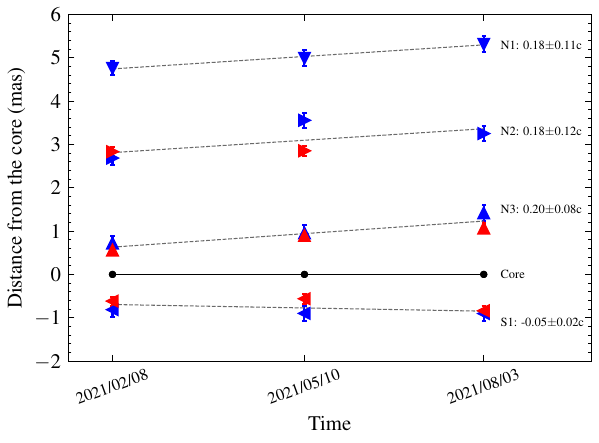}
\end{center}
\vspace{-0.5cm}
\caption{Radial distance from the core versus time for the jet components in NGC\,4261 (left) and M\,104 (right) (Yan et al. \cite{Yan2023,Yan2024}).
\label{fig:jet motions}}
\end{figure*}

\section{Results and Discussion}
\subsection{Jet kinematics}
Using multi-epoch 15\,GHz data for NGC\,4261 and a multi-epoch 12/22\,GHz dataset for M\,104, we have measured jet motions in these two LLAGNs on sub-parsec scales. As shown in Fig.\ref{fig:jet motions}, the apparent speeds in the approaching jet of NGC\,4261 range from  $0.31\pm0.14\,c$ to $0.59\pm0.40\,c$, while the speed in the counter-jet is $0.32\pm0.14\,c$. For M\,104, the measured apparent speeds are about $0.20\pm0.08\,c$ for the approaching jet and $0.05\pm0.02\,c$ for the receding jet.

Previously, Piner et al. (\cite{Piner2001}) reported an apparent speed of $0.83\pm0.11$ mas/yr ($1\,c$ corresponds to $2\,\text{mas/yr}$) at about 5--6 mas from the core based on two-epoch observations. We found that the apparent speeds measured in this study are consistent with the previous results.

\begin{figure*}[htbp!]
\begin{center}
        \includegraphics[width=0.4\textwidth]{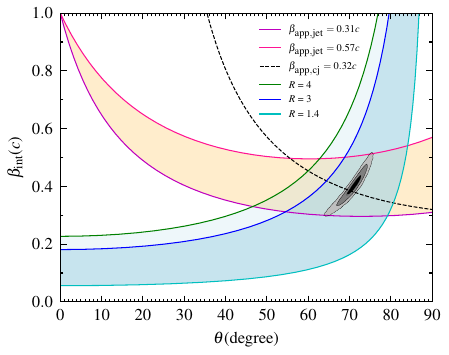}
        \hspace{5mm}
        \includegraphics[width=0.41\textwidth]{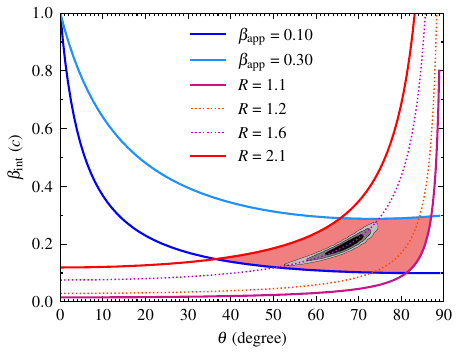}
\end{center}
\vspace{-0.5cm}
\caption{Possible range of the viewing angle and intrinsic velocity of NGC\,4261 (right) and M\,104 (left) jets. The contours represent the joint probability distribution of the viewing angle and intrinsic speed (Yan et al. \cite{Yan2023,Yan2024}).
\label{fig:jet VA}}
\end{figure*}

\begin{figure*}
\begin{center}
        \hspace{-0.7cm}
        \includegraphics[width=0.41\textwidth]{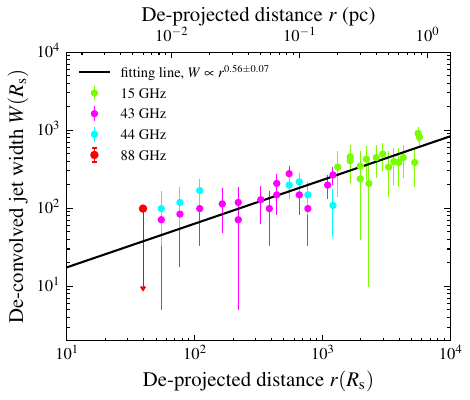}
        \hspace{1.3cm}
        \includegraphics[width=0.33\textwidth]{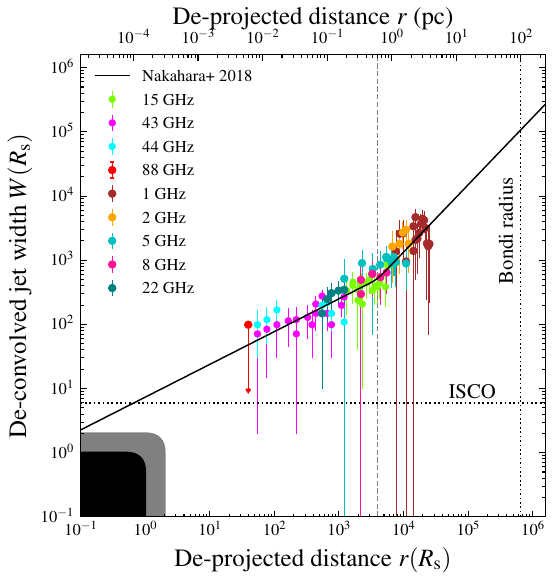}
\end{center}
\vspace{-0.5cm}
\caption{The jet collimation profile of NGC\,4261 (Yan et al. \cite{Yan2023}).
\label{fig:NGC4261 collimation}}
\end{figure*}

\subsection{Jet viewing angle}
The presence of the counter-jet allows us to measure the jet-to-counter-jet brightness ratio on the studied scales, which ranges from 1.4 to 3 for NGC\,4261 and from 1.1 to 2.1 for M\,104\footnote{For M\,104, we have also included the measurements of Hada et al. \cite{Hada2013}.}. Assuming that the jets in NGC\,4261 and M\,104 are intrinsically symmetric, with the same brightness and speed, we can use the determined brightness ratio and apparent speed to constrain the viewing angle ($\theta$) and the intrinsic speed ($\beta_{\rm int}$) of both jets. The results are shown in Fig.\ref{fig:jet VA}. On sub-parsec scales, we obtained an intrinsic speed range from $\sim$ 0.30\,$c$ to 0.55\,$c$ for NGC\,4261 and from $\sim0.10\,c$ to $0.40\,c$ for M\,104. The viewing angle range was derived to be $54^{\circ} \sim 84^{\circ}$ for NGC\,4261 and  $\theta \ga 37^{\circ}$ for M\,104, with the most probable values being ${71^{\circ}}\pm2^{\circ} (1\sigma)$ (or ${71^{\circ}}^{+4^{\circ}}_{-9^{\circ}} (3\sigma)$) and ${66^{\circ}}^{+4^\circ}_{-6^\circ}(1\sigma)$, respectively, which are based on the joint probability distribution of $\theta$ and $\beta_{\rm int}$.

\vspace{-0.4cm}
In Piner et al. (\cite{Piner2001}), the jet viewing angle for NGC\,4261 was measured to be $63^{\circ}\pm3^{\circ}$, which aligns with our constrained range. However, for M\,104, Hada et al. (\cite{Hada2013}) reported a jet viewing angle of $\theta\la25^{\circ}$, whereas we measured a larger range of $\theta \ga 37^{\circ}$. Several pieces of observational evidence support this larger viewing angle, including the nearly symmetric two-sided jets detected by VLBA at low frequencies, as well as the nearly edge-on torus observed in optical and near-infrared observations (see the discussion in Yan et al. \cite{Yan2024}).

\subsection{Jet collimation profile of NGC\,4261}
Using the high-resolution data, we measured the collimation profile of the innermost jet in NGC\,4261, with the form $W \propto r^{0.56\pm0.07}$, where $W$ is the deconvolved jet width and $r$ is the de-projected distance from the black hole (see the left panel in Fig.\ref{fig:NGC4261 collimation}). This corresponds to a parabolic jet shape. We also measured the width of the downstream jet using the previous multi-frequency VLBA data of Haga et al. \cite{Haga2015}. As shown in the right panel of Fig.\ref{fig:NGC4261 collimation}, these results clearly indicate a transition from parabolic to conical shape of the jet collimation profile. 

We emphasize that the jet collimation is already completed on sub-parsec scales, which is significantly smaller than the Bondi radius. Two possible reasons have been proposed: the external pressure provided by the ADAF and/or the disk wind, and the low initial magnetization of the jet (see the discussion in Yan et al. \cite{Yan2023}).

\begin{figure*}[htbp!]
\begin{center}
        \hspace{-0.35cm}
        \includegraphics[width=0.4\textwidth]{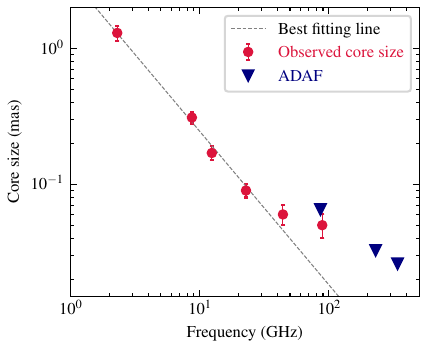}
        \hspace{0.5cm}
        \includegraphics[width=0.48\textwidth]{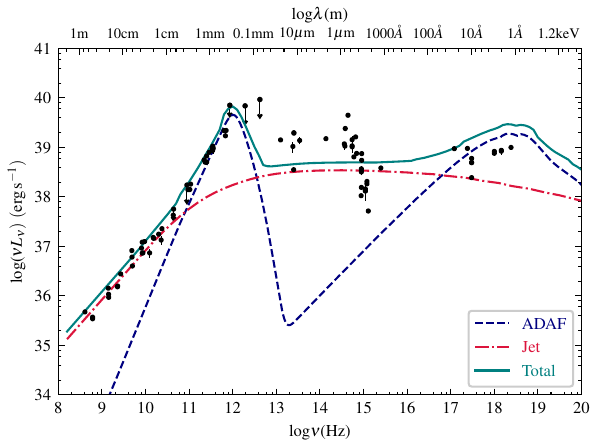}
\end{center}
\vspace{-0.5cm}
\caption{Frequency dependence of core size (left) and modeled broad-band SED (right) for M\,104 (Yan et al. \cite{Yan2024}).
\label{fig:M104 core size}}
\vspace{-0.6cm}
\end{figure*}

\begin{figure}
\begin{center}
        \includegraphics[width=0.4\textwidth]{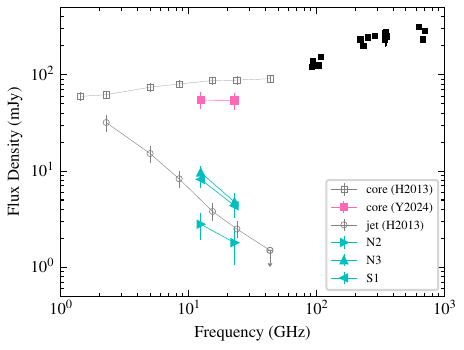}
\end{center}
\vspace{-0.4cm}
\caption{The radio spectra of M\,104.
\label{fig:M104 spectra}}
\vspace{-0.5cm}
\end{figure}

\subsection{Possible origin of the millimeter emission in M\,104} \label{subsec: origin of mm emission}
In Fig.\ref{fig:M104 core size} (left), we show the frequency dependence of the VLBI core size ($d$) in M\,104. The dependence is best described by a power-law of $d\propto\nu^{-1.13\pm0.04}$ at lower frequencies, which is in full agreement with previous results by Hada et al. \cite{Hada2013} ($d \propto\nu^{-1.20\pm0.08}$). At higher frequencies (i.e., 88\,GHz), however, the core size seems to deviate from this relationship. 

The combination of the above results with the spectral energy distribution (SED) of M\,104 may help explain the flattening of the frequency-size dependence in the millimeter regime. Fig.\ref{fig:M104 core size} (right) shows the broadband SED of M\,104, along with the fitting results based on a coupled Advection Dominated Accretion Flow (ADAF)-jet model developed by Xie et al. (\cite{Xie2016}). As shown, the radio emission in M\,104 is predominantly jet-dominated at frequencies below about 100\,GHz. Beyond this frequency range, however, the contribution from the accretion flow becomes increasingly significant. This implies that the observed mm-VLBI core may be  primarily dominated by emission from the accretion flow rather than the jet, potentially explaining the flattening of the frequency-size dependence in the millimeter regime.

Therefore, we derived the expected ADAF sizes at 86, 230, and 340\,GHz using this model, which are also shown in the left panel of Fig.\ref{fig:M104 core size}. As can be seen, all the predicted sizes are significantly larger than the extrapolated core sizes from lower frequencies. Interestingly, the predicted ADAF size at 86\,GHz agrees well with the measured core size at 88\,GHz, possibly indicating the dominance of the accretion flow over the jet. This is similar to the recent findings in M\,87, where 86\,GHz observations showed that the VLBI core is spatially resolved into a ring-like structure that is primarily dominated by the emission from the accretion flow (Lu et al. \cite{Lu2023}).

\subsection{Spectra of M\,104}
Using the quasi-simultaneously observed 12/22\,GHz dataset, we briefly examined the spectral properties of the core and the jet components. For the core, the averaged flux densities at 12 and 22\,GHz are quite close (see Fig.\ref{fig:M104 spectra}), indicating a flat spectrum. Conversely, both the approaching jet (N2 and N3, see Fig.\ref{fig:jet motions}) and the receding jet (S1) show a steep spectrum, with an averaged spectral index value of $\alpha = -0.97 \pm 0.07$ (see Fig.\ref{fig:M104 spectra}).  These results are consistent with those obtained by a multi-frequency analysis by Hada et al. \cite{Hada2013} (see Fig.\ref{fig:M104 spectra}). Notably, the counter-jet shows an optically thin synchrotron spectrum, indicating that free-free absorption processes are not significant at the observing frequencies and the studied physical scales.

Additionally, we included archival sub-millimeter data from the Atacama Large Millimeter/submillimeter Array (ALMA; represented by the black squares in Fig.\ref{fig:M104 spectra}). These data reveal a ``sub-millimeter bump" (see also Fig.\ref{fig:M104 core size}, right), although it may be partially contaminated by thermal emission from dust in the nucleus. As discussed in Sec.\ref{subsec: origin of mm emission}, this sub-millimeter bump could be attributed to the synchrotron radiation from thermal electrons in the ADAF.

\section{Conclusions}

\begin{itemize}

    \item[(1)] The sub-parsec jets in both NGC\,4261 and M\,104 were found to be sub-relativistic, with jet viewing angle being ${71^{\circ}}\pm2^{\circ} (1\sigma)$ (or ${71^{\circ}}^{+4^{\circ}}_{-9^{\circ}} (3\sigma)$) and ${66^{\circ}}^{+4^\circ}_{-6^\circ}(1\sigma)$, respectively. 
    
    \item[(2)] The jet collimation in NGC\,4261 has already been completed on sub-parsec scales, which may be ascribed to the confinement of the jet by the external pressure and/or the low initial magnetization of the jet.
    
    \item[(3)] The origin of the millimeter emission and the sub-millimeter bump in M\,104 can be explained by the accretion flow, rather than the jet. 
    
\end{itemize}

\end{document}